\documentclass{article}%
\usepackage{graphicx}
\usepackage{amsmath}%
\usepackage{amsfonts}%
\usepackage{amssymb}
\newtheorem{theorem}{Theorem}
\newtheorem{acknowledgement}[theorem]{Acknowledgement}

\begin{document}

\title{{\Large Symmetries in Constrained Systems}}
\author{D.M. Gitman\thanks{Institute of Physics, University of S\~{a}o Paulo, Brazil;
e-mail: gitman@dfn.if.usp.br} \ and I.V. Tyutin\thanks{Lebedev Physics
Institute, Moscow, Russia; e-mail: tyutin@lpi.ru}}
\date{\today }
\maketitle

\begin{abstract}
We describe symmetry structure of a general singular theory
(theory with constraints in the Hamiltonian formulation), and, in
particular, we relate the structure of gauge transformations with
the constraint structure. We show that any symmetry transformation
can be represented as a sum of three kinds of symmetries: global,
gauge, and trivial symmetries. We construct explicitly all the
corresponding conserved charges as decompositions in a special
constraint basis. The global part of a symmetry does not vanish on
the extremals, and the corresponding charge does not vanish on the
extremals as well. The gauge part of a symmetry does not vanish on
the extremals, but the gauge charge vanishes on them. We stress
that the gauge charge necessarily contains a part that vanishes
linearly in the first-class constraints and the remaining part of
the gauge charge vanishes quadratically on the extremals. The
trivial part of any symmetry vanishes on the extremals, and the
corresponding charge vanishes quadratically on the extremals.
\end{abstract}

\section{Introduction}

Our aim is to study the symmetry structure of a general singular theory, and,
in particular, to relate this structure to the constraint structure in the
Hamiltonian formulation. For simplicity, we consider finite-dimensional models
whose actions are of the form%
\begin{align}
&  S\left[  q\right]  =\int L\left(  q,\dot{q}\right)  \,dt\,,\;q=\left(
q^{a},\;a=1,...,n\right)  \,,\nonumber\\
&  \frac{\delta S}{\delta q^{a}}=\frac{\partial L}{\partial q^{a}}%
-\frac{d}{dt}\frac{\partial L}{\partial\dot{q}^{a}}%
=0-\mathrm{Euler-Lagrange\;equations}, \label{I.1a}%
\end{align}
where $L\left(  q,\dot{q}\right)  $ is a Lagrange function. All such theories
can be divided into two classes according to the Hessian's value,%
\begin{equation}
\mathrm{Hessian}=\det\frac{\partial^{2}L}{\partial\dot{q}^{a}\partial\dot
{q}^{b}}=\left\{
\begin{array}
[c]{c}%
\neq0\;\mathrm{nonsingular\;theory}\\
=0\;\mathrm{singular\;theory}%
\end{array}
\right.  . \label{I.2}%
\end{equation}
Singular Lagrangian theories are theories with constraints in the Hamiltonian
formulation \cite{Singular}. In particular, theories with first-class
constraints (FCC) are gauge theories.

A finite transformation $q\left(  t\right)  \rightarrow q^{\prime}\left(
t\right)  $ is a symmetry of $S$ if%
\begin{equation}
L\left(  q,\dot{q}\right)  \rightarrow L^{\prime}\left(  q,\dot{q}\right)
=L\left(  q,\dot{q}\right)  +\frac{dF}{dt}\,,\label{I.5}%
\end{equation}
where $F$ is a local function (such transformations are called N\"{o}ether
symmetries). The finite symmetry transformations can be discrete, continuous
global, gauge, and trivial. Continuous global symmetry transformations are
parametrized by a set of time-independent parameters $\nu^{\alpha}$%
,\ $\alpha=1,...,r$. The infinitesimal form of a continuous global symmetry
transformation reads $\delta q^{a}(t)=\rho_{\alpha}^{a}(t)\nu^{\alpha},$ where
$\rho_{\alpha}^{a}(t)$ are generators of the global symmetry transformations.
Continuous symmetry transformations are gauge transformations (or local
symmetry transformations) if they are parametrized by arbitrary functions of
time, the gauge parameters (in the case of field theory, the gauge parameters
depend on all space-time variables). The infinitesimal form of a gauge
transformation reads
\begin{equation}
\delta q^{a}(t)=\hat{\Re}_{\alpha}^{a}\left(  t\right)  \nu^{\alpha
}(t),\label{I.7}%
\end{equation}
where $\nu^{\alpha}\left(  t\right)  $, $\alpha=1,...,r$ are time-dependent
gauge parameters. The quantities $\hat{\Re}_{\alpha}^{a}\left(  t\right)  $
are generators of gauge transformations. Under some natural suppositions about
the structure of the Lagrange function, \textbf{one can prove that }$\hat{\Re
}_{\alpha}^{a}\left(  t\right)  $\textbf{\ are local operators} \cite{197}.
The existence of infinitesimal gauge transformations with generators implies
the existence of the corresponding gauge identities, which present identities
between the Euler-Lagrange equations.

\textbf{For any action there exist trivial symmetry transformations,}
\begin{equation}
\delta_{\mathrm{tr}}q^{a}=\hat{U}^{ab}\frac{\delta S}{\delta q^{b}}\,,
\label{I.9}%
\end{equation}
\textbf{where }$\hat{U}$\textbf{\ is an antisymmetric local operator}, that is
$\left(  \hat{U}^{T}\right)  ^{ab}=-\hat{U}^{ab}$ . The trivial symmetry
transformations do not affect genuine trajectories. Two symmetry
transformations $\delta_{1}q$ and $\delta_{2}q$ are called equivalent
($\delta_{1}q\sim\delta_{2}q$)\ whenever they differ by a trivial symmetry
transformation,
\begin{equation}
\delta_{1}q\sim\delta_{2}q\Longleftrightarrow\delta_{1}q-\delta_{2}%
q=\delta_{\mathrm{tr}}q\,. \label{I.10}%
\end{equation}
Thus, all the symmetry transformations of an action $S$ can be divided
into\emph{\ }equivalence classes.

\textbf{Any symmetry transformation implies a conservation law (}N\"{o}ether
theorem)\textbf{:}%
\begin{align}
&  \frac{dG}{dt}=-\delta q^{a}\frac{\delta S}{\delta q^{a}}=O\left(
\frac{\delta S}{\delta q}\right)  \Longrightarrow
G=\mathrm{const.\;on\;extremals,}\label{I.11}\\
&  G=P-F\,,\;P=\frac{\partial L}{\partial\dot{q}^{a}}\delta q^{a},\;\delta
L=\frac{dF}{dt}\,.\nonumber
\end{align}
The local function $G$ is referred to as the conserved charge\emph{\ }related
to the symmetry $\delta q$ of the action $S$. The quantities $\delta q,\;S,$
and $G$ are related by the equation (\ref{I.11}). In what follows, we call
this equation the symmetry equation.

\textbf{Any gauge symmetry generates a conserved charge }$G$\textbf{\ which
depends locally on gauge parameters and on their time-derivatives, and
vanishes on the extremals}\footnote{For us, extremals are local functions
$\delta S/\delta q$ and any linear combinations of these functions and their
time derivatives.}\textbf{\ }(the latter fact was already familiar to
N\"{o}ether)%
\begin{equation}
G=O\left(  \frac{\delta S}{\delta q}\right)  \,. \label{I.12}%
\end{equation}

An important inverse statement holds true. Namely: \textbf{If a global
symmetry transformation generates a conserved charge that vanishes on the
extremals then the corresponding action obeys a gauge symmetry. At the same
time the initial global symmetry is a reduction of the corresponding gauge
symmetry to constant values of the gauge parameters }\cite{FulGiT}\textbf{.}

At present almost all modern physical models are formulated as gauge theories.
Thus, the study of the general gauge theory is an important mathematical and
physical problem. In particular, the following questions are of especial interest:

\textbf{How many gauge transformations (with independent gauge parameters) are
there for a given action?}

\textbf{What is the structure of the gauge generators (how many time
derivatives they contain) for a given action? What is the structure of an
arbitrary symmetry of the action of a singular theory? }

\textbf{Is there a constructive procedure to find all the gauge
transformations for a given action?}

\textbf{How can one relate the constraint structure in the Hamiltonian
formulation with the symmetry structure of the Lagrangian action?}

These problems were partially considered in the works
\cite{GitTy90,HenTe92,BorTy98}. In this talk, we represent the recent progress
in attempts to answer the above questions.

\section{Symmetry equation and orthogonal constraint basis}

For the study of the symmetry structure, we start with the consideration of
the Hamiltonian action $S_{\mathrm{H}}$ (there exists an isomorphism between
symmetry classes of the Lagrangian $S$ and the Hamiltonian $S_{\mathrm{H}}$
actions).%
\begin{align*}
&  S\left[  q\right]  =\int L\left(  q,\dot{q}\right)  \,dt\Longleftrightarrow
S_{\mathrm{H}}\left[  \mathbf{\eta}\right]  =\int\left[  p\dot{q}-H^{\left(
1\right)  }\left(  \mathbf{\eta}\right)  \right]  dt\,,\;\mathbf{\eta}=\left(
\eta,\lambda\right)  \,,\;\eta=\left(  q,p\right)  \,,\\
&  H^{\left(  1\right)  }\left(  \mathbf{\eta}\right)  =H\left(  \eta\right)
+\lambda\Phi^{\left(  1\right)  }\left(  \eta\right)  \,;\\
&  \frac{\delta S_{\mathrm{H}}}{\delta\mathbf{\eta}}=0\Longrightarrow\left\{
\begin{array}
[c]{c}%
\dot{\eta}=\{\eta,H^{\left(  1\right)  }\}\,,\\
\Phi^{\left(  1\right)  }\left(  \eta\right)  =0\,,
\end{array}
\right.
\end{align*}
where $\Phi^{\left(  1\right)  }\left(  \eta\right)  $ are
primary\ constraints, $\eta=\left(  q,p\right)  $ are phase-space variables,
and $\lambda$ are Lagrange multipliers to primary constraints.

One can see that if $\delta\mathbf{\eta=}\left(  \delta q,\delta
p,\delta\lambda\right)  $ is a symmetry of the Hamiltonian action
$S_{\mathrm{H}}$, then $\delta_{L}q$ is a symmetry of the Lagrangian action
$S,$%
\[
\delta_{L}q=\left.  \delta q\right|  _{p(q,\dot{q}),\,\lambda(q,\dot{q})}\,.
\]
The symmetry equation for the action $S_{\mathrm{H}}$ reads%
\begin{equation}
\delta\mathbf{\eta}\frac{\delta S_{\mathrm{H}}}{\delta\mathbf{\eta}}%
+\frac{dG}{dt}=0\,, \label{symeq}%
\end{equation}
where $G$ is the conserved charge. The charge $G$ and all the variations
depend on all the variables and their time derivatives locally. One can study
the symmetry of an action by solving the symmetry equation.

It turns out that the symmetry equation can be easily analyzed (solved) by
algebraic methods if one chooses the so called orthogonal constraint basis. In
the work \cite{196}, we have demonstrated that there exists a constraint
reorganization of the first-class constraints (FCC) and of the second-class
constraints (SCC) consistent with the Dirac procedure, i.e., the
reorganization does not violate the decomposition of the
constraints\ according to their stages in the Dirac procedure. Namely:

\textbf{It is possible to reorganize the independent constraints }$\Phi
$\ \textbf{obtained in the Dirac procedure such that: the complete set of
constraints is divided into SCC }$\varphi$ \textbf{and FCC }$\chi$\textbf{. At
the same time, it is decomposed into groups according to the stages of the
Dirac procedure,}
\begin{align*}
&  \Phi=\left(  \varphi,\,\chi\right)  =\left(  \Phi^{\left(  i\right)
}\right)  \,,\;i=1,...,\aleph\,,\\
&  \Phi^{\left(  i\right)  }=(\varphi^{(i)};\chi^{\left(  i\right)
}),\;\varphi=(\varphi^{(i)})\,,\;\chi=(\chi^{\left(  i\right)  })\,.
\end{align*}
\ \textbf{Here }$\Phi^{\left(  i\right)  }$\textbf{\ are constraints of the
}$i$\textbf{-th stage, }$\varphi^{(i)}$\textbf{\ are SCC of the }%
$i$\textbf{-th stage, }$\chi^{\left(  i\right)  }$\textbf{\ are FCC of the
}$i$\textbf{-th stage, and }$\aleph$\textbf{\ is the number of stages of the
Dirac procedure. It may turn out that after a certain stage new independent
FCC (SCC) do not appear anymore. We are going to denote this stages by
}$\aleph_{\chi}$ \textbf{(}$\aleph_{\varphi}$)\textbf{. Obviously, }%
$\aleph=\max(\aleph_{\chi},\aleph_{\varphi})$\textbf{. In addition, the
constraints in each stage are divided into groups,}
\begin{align}
\varphi^{\left(  i\right)  }  &  =\left(  \varphi^{\left(  i|s\right)
}\right)  ,\;s=i,...,\aleph_{\varphi}\,;\nonumber\\
\chi^{\left(  i\right)  }  &  =\left(  \chi^{\left(  i|a\right)  }\right)
,\;a=i,...,\aleph_{\chi}\,. \label{s.1}%
\end{align}
\textbf{Such a division creates chains of constraints. Thus, there exist
}$\aleph_{\varphi}$\textbf{\ chains of SCC}
\[
\varphi^{\left(  ...|s\right)  }=\left(  \varphi^{\left(  i|s\right)
},\;i=1,...,s\right)  \,,\;s=1,...,\aleph_{\varphi}\,,
\]
\textbf{labeled by the index} $s,$ \textbf{and }$\aleph_{\chi}$%
\textbf{\ chains of FCC}
\[
\chi^{\left(  ...|a\right)  }=\left(  \chi^{\left(  i|a\right)  }%
,\;i=1,...,a\right)  \,,\;a=1,...,\aleph_{\chi}\,
\]
\textbf{\ labeled by the index} $a.$ \textbf{Within the Dirac procedure, the
group }$\varphi^{\left(  1|s\right)  }$ \textbf{of primary SCC \ produces SCC
of the second stage, third stage, and so on, which belong to the same
chain,\ }$\varphi^{\left(  1|s\right)  }\rightarrow\varphi^{\left(
2|s\right)  }\rightarrow\varphi^{\left(  3|s\right)  }\rightarrow\cdot
\cdot\cdot\rightarrow\varphi^{\left(  s|s\right)  }.$\textbf{\ The chain of
SCC labeled by the number }$s$\textbf{\ ends with the group of the }%
$s$\textbf{-th-stage constraints. The consistency conditions for the latter
group determine\ the Lagrange multipliers }$\lambda_{\varphi}$\textbf{\ to be
}$\bar{\lambda}\,.$ \textbf{At the same time, the group }$\chi^{\left(
1|a\right)  }$\textbf{\ of primary FCC\ produces FCC of the second stage,
third stage, and so on, which belong to the same chain,}$\;\chi^{\left(
1|a\right)  }\rightarrow\chi^{\left(  2|a\right)  }\rightarrow\chi^{\left(
3|a\right)  }\rightarrow\cdot\cdot\cdot\rightarrow\chi^{\left(  a|a\right)  }%
$\textbf{. We call such organized set of constraints the orthogonal constraint
basis. The described hierarchy of constraints in the orthogonal basis (in the
Dirac procedure) looks schematically as follows:}
\[%
\begin{array}
[c]{lllllllllll}%
\varphi^{\left(  1|1\right)  } & \rightarrow & \bar{\lambda}_{1} &  &  &  &  &
&  &  & \\
\varphi^{\left(  1|2\right)  } & \rightarrow & \varphi^{\left(  2|2\right)  }
& \rightarrow & \bar{\lambda}_{2} &  &  &  &  &  & \\
\vdots & \rightarrow & \vdots & \rightarrow & \vdots & \vdots & \vdots &  &  &
& \\
\varphi^{\left(  1|\aleph-1\right)  } & \rightarrow & \varphi^{\left(
2|\aleph-1\right)  } & \rightarrow & \varphi^{\left(  3|\aleph-1\right)  } &
\cdots & \varphi^{\left(  \aleph-1|\aleph-1\right)  } & \rightarrow &
\bar{\lambda}_{\aleph-1} &  & \\
\varphi^{\left(  1|\aleph\right)  } & \rightarrow & \varphi^{\left(
2|\aleph\right)  } & \rightarrow & \varphi^{\left(  3|\aleph\right)  } &
\cdots & \varphi^{\left(  \aleph-1|\aleph\right)  } & \rightarrow &
\varphi^{\left(  \aleph|\aleph\right)  } & \rightarrow & \bar{\lambda}%
_{\aleph}\\
\chi^{\left(  1|\aleph\right)  } & \rightarrow & \chi^{\left(  2|\aleph
\right)  } & \rightarrow & \chi^{\left(  3|\aleph\right)  } & \cdots &
\chi^{\left(  \aleph-1|\aleph\right)  } & \rightarrow & \chi^{\left(
\aleph|\aleph\right)  } & \rightarrow & O(\Phi)\\
\chi^{\left(  1|\aleph-1\right)  } & \rightarrow & \chi^{\left(
2|\aleph-1\right)  } & \rightarrow & \chi^{\left(  3|\aleph-1\right)  } &
\cdots & \chi^{\left(  \aleph-1|\aleph-1\right)  } & \rightarrow &
O(\Phi^{\left(  ...\aleph-1\right)  }) &  & \\
\vdots & \rightarrow & \vdots & \rightarrow & \vdots & \vdots & \vdots &  &  &
& \\
\chi^{\left(  1|2\right)  } & \rightarrow & \chi^{\left(  2|2\right)  } &
\rightarrow & O(\Phi^{\left(  ...2\right)  }) &  &  &  &  &  & \\
\chi^{\left(  1|1\right)  } & \rightarrow & O(\Phi^{\left(  1\right)  }) &  &
&  &  &  &  &  &
\end{array}
.
\]
\textbf{\ The chain of FCC labeled by the number }$a$\textbf{\ ends with the
group of the }$a$\textbf{-th-stage constraints. Their consistency conditions
do not determine\ any multipliers and any new constraints. The Lagrange
multipliers }$\lambda_{\chi}$\textbf{\ are not determined by the Dirac
procedure (and by the complete set of equations of motion). Thus, all the
constraints in a chain are of the same class. One ought to say that the
numbers of constraints in each stage in the same chain are the same. At the
same time, each chain may be either empty or contain several functions. Thus,
whenever FCC (SCC) exist, the corresponding primary FCC (SCC) do exist.}

\textbf{The Poisson brackets of SCC from different chains of the orthogonal
basis vanish on the constraint surface}%
\[
\left\{  \varphi^{\left(  i|s\right)  },\varphi^{\left(  j|v\right)
}\right\}  =O\left(  \Phi\right)  \,,\;s\neq v\,.
\]
\textbf{In addition,}
\begin{align*}
&  \left\{  \varphi^{\left(  i|s\right)  },H^{\left(  1\right)  }\right\}
=\varphi^{\left(  i+1|s\right)  }+O\left(  \Phi^{\left(  1\right)  }%
,...,\Phi^{\left(  i\right)  }\right)  \,,\;i=1,...,\aleph_{\varphi
}-1,\;s=i+1,...,\aleph_{\varphi}\,,\\
&  \left\{  \varphi^{\left(  1|s\right)  },\varphi^{\left(  s|s\right)
}\right\}  =\theta\,,\;\det\theta^{s}\neq0\,;\\
&  \left\{  \chi^{\left(  i|a\right)  },H^{\left(  1\right)  }\right\}
=\chi^{\left(  i+1|a\right)  }+O\left(  \Phi^{\left(  1\right)  }%
,...,\Phi^{\left(  i\right)  }\right)  \,,\;i=1,...,\aleph_{\chi
}-1,\;a=i+1,...,\aleph_{\chi}\,,\\
&  \left\{  \chi^{\left(  a|a\right)  },H^{\left(  1\right)  }\right\}
=O\left(  \Phi^{\left(  1\right)  },...,\Phi^{\left(  a\right)  }\right)  \,.
\end{align*}
\textbf{The consistency conditions for SCC }$\varphi^{\left(  i|i\right)  }%
$\textbf{ of the }$i$\textbf{-th stage}%
\[
\left\{  \varphi^{\left(  s|s\right)  },H^{\left(  1\right)  }\right\}  =0
\]
\textbf{ allows one to determine }$\lambda_{\varphi}^{s}$\textbf{ multipliers}.

We stress, \textbf{that the consistency conditions for SCC }$\varphi^{\left(
i|s\right)  },$\textbf{\ }$s>i$\textbf{ of the }$i$\textbf{-th stage produce
SCC }$\varphi^{\left(  i+1|s\right)  }$\textbf{ of the }$i+1$\textbf{-the
stage. The consistency conditions for FCC }$\chi^{\left(  i|s\right)  }%
,$\textbf{\ }$s>i$\textbf{ of the }$i$\textbf{-th stage produce FCC }%
$\chi^{\left(  i+1|s\right)  }$\textbf{ of the }$i+1$\textbf{-the stage. The
consistency conditions for FCC }$\chi^{\left(  i|i\right)  }$\textbf{ of the
}$i$\textbf{-th stage do not produce any new constraints and do not determine
any Lagrange multipliers.}

Such properties of the constraint basis are extremely helpful for analyzing
the symmetry equation. In particular, they allow one to guess (and then to
strictly prove) the form of the conserved charges as decompositions in the
orthogonal constraint basis. For example, these properties imply that SCC
$\varphi^{\left(  i|i\right)  }$ cannot enter linearly into the conserved
charges. At the same time, one can see that only FCC $\chi^{\left(
i|i\right)  }$ enter the gauge charges multiplied by independent gauge
parameters, other FCC $\chi^{\left(  i|a\right)  },\;a>i$ are multiplied by
factors that must contain derivatives of the same gauge parameters.

\section{What can be proved solving the symmetry equation in orthogonal
constraint basis?}

\textbf{I.} \textbf{For any theory (singular or non-singular) any symmetry
transformations that vanish on the equations of motion are trivial.}

\textbf{II.} \textbf{In theories with FCC there exist nontrivial symmetries
}$\delta_{\nu}\mathbf{\eta}$, $G_{\nu}$ \textbf{of the Hamiltonian action
}$S_{\mathrm{H}}$ \textbf{that are} \textbf{gauge transformations. These
symmetries are parametrized by the gauge parameters }$\nu_{i}$.\textbf{\ The
latter parameters are arbitrary functions of time }$t$\textbf{.}

\textbf{The number of the gauge parameters }$\left[  \nu\right]  $ \textbf{is
equal to the number of the primary FCC }$\left[  \chi^{\left(  1\right)
}\right]  $\textbf{,}%
\[
\left[  \nu\right]  =\left[  \chi^{\left(  1\right)  }\right]  \,.
\]

\textbf{The corresponding conserved charge (the gauge charge) is a local
function }$G_{\nu}=G_{\nu}\left(  \eta,\lambda^{\left[  l\right]  }%
,\nu^{\lbrack l]}\right)  $, \textbf{which vanishes on the extremals. The
gauge charge has the following decomposition with respect to the orthogonal
constraint basis:}%
\begin{equation}
G_{\nu}=\sum_{i=1}^{\aleph_{\chi}}\nu_{i}\chi^{\left(  i|i\right)  }%
+\sum_{i=1}^{\aleph_{\chi}-1}\sum_{a=i+1}^{\aleph_{\chi}}C_{i|a}^{\chi}%
\chi^{\left(  i|a\right)  }+\sum_{i=1}^{\aleph_{\chi}-1}\sum_{s=i+1}^{\aleph
}C_{i|s}^{\varphi}\varphi^{\left(  i|s\right)  }\,.\label{sdc.4}%
\end{equation}
\textbf{Here }$C_{i|s}^{\varphi}\left(  \eta,\lambda^{\left[  l\right]  }%
,\nu^{\lbrack l]}\right)  $\textbf{\ and }$C_{i|a}^{\chi}\left(  \eta
,\lambda^{\left[  l\right]  },\nu^{\lbrack l]}\right)  $\textbf{\ are some
local functions,} \textbf{which are determined by the symmetry equation in an
algebraic way. It turns out that }$C_{i|s}^{\varphi}=O\left(  I\right)
,$\textbf{\ where }$I=\delta S_{\mathrm{H}}/\delta\mathbf{\eta}$\textbf{\ are
extremals. The gauge charge depends both on the gauge parameters and on their
time derivatives up to a finite} \textbf{order. Namely,}%
\begin{equation}
G_{\nu}=\sum_{i=1}^{\aleph_{\chi}}\sum_{m=0}^{i-1}G_{m}^{i}(\eta
,\lambda^{\lbrack l]})\nu_{i}^{[m]}\,.\label{sdc.5}%
\end{equation}
\textbf{where }$G_{m}^{i}(\eta,\lambda^{\lbrack l]})$\textbf{\ are some local
functions.} \textbf{The total number of independent gauge parameters together
with their time derivatives, that enter essentially in the gauge charge}
\textbf{is equal to the number of all FCC }$\left[  \chi\right]  $\textbf{,}%
\[
\sum_{m=0}[\nu^{\lbrack m]}]=\left[  \chi\right]  \,.
\]
\textbf{The gauge charge} \textbf{is the generating function for the
variations }$\delta\eta$\textbf{\ of the phase-space variables,}%
\begin{equation}
\delta_{\nu}\eta=\left\{  \eta,G_{\nu}\right\}  =\left\{  \eta,\eta
^{A}\right\}  \frac{\partial G_{\nu}}{\partial\eta^{A}}\,.\label{sdc.7}%
\end{equation}
(\textbf{Note that here the Poisson bracket acts only on the explicit
dependence on }$\eta$ \textbf{of the gauge charge}.) \textbf{ The variations
}$\delta_{\nu}\lambda$\textbf{\ contain additional time derivatives of the
gauge parameters, namely, they have the form}%
\begin{equation}
\delta_{\nu}\lambda=\sum_{i=1}^{\aleph_{\chi}}\sum_{m=0}^{i}\Upsilon_{m}%
^{i}(\eta,\lambda^{\lbrack l]})\,\nu_{i}^{[m]}\,,\label{sdc.8}%
\end{equation}
\textbf{where }$\Upsilon_{m}^{i}$\textbf{\ are some local functions,}
\textbf{which can be determined from the symmetry equation in an algebraic way.}

Thus, the gauge charge $G_{\nu}$ have the following structure%
\begin{equation}
G_{\nu}=\sum_{m=1}^{\aleph_{\chi}}\sum_{b=m}^{\aleph_{\chi}}G^{mb}%
(\eta,\lambda^{\lbrack l]})\nu_{b}^{[m-1]}\,,\label{sdc.19}%
\end{equation}
where the local functions $G^{mb}(\eta,\lambda^{\lbrack l]})$ have the form%
\begin{equation}
G^{mb}=\sum_{k=1}^{\aleph_{\chi}}\sum_{a=k}^{\aleph_{\chi}}\chi^{\left(
k|a\right)  }\mathcal{C}_{ka}^{mb}(\eta,\lambda^{\lbrack l]})+O\left(
I^{2}\right)  \,,\label{sdc.20}%
\end{equation}
and $\mathcal{C}_{ka}^{mb}(\eta,\lambda^{\lbrack l]})$ are some local
functions. Thus,%
\begin{equation}
G_{\nu}=O\left(  \chi\right)  +O\left(  I^{2}\right)  \,.\label{sdc.20a}%
\end{equation}

The form of the variations $\delta_{\nu}\eta$ follows from (\ref{sdb.7}),%
\begin{equation}
\delta_{\nu}\eta=\left(  \sum_{k=1}^{\aleph_{\chi}}\sum_{a=k}^{\aleph_{\chi}%
}\right)  \left(  \sum_{m=1}^{\aleph_{\chi}}\sum_{b=m}^{\aleph_{\chi}}\right)
\left\{  \eta,\chi^{\left(  k|a\right)  }\right\}  \mathcal{C}_{ka}^{mb}%
\nu_{b}^{[m-1]}+O\left(  I\right)  \,. \label{sdc.21}%
\end{equation}

After the gauge charge has been determined, the variations $\delta_{\nu
}\lambda$ can be found from the equation (\ref{sdc.9}). Their general
structure is given by Eq. (\ref{sdc.8}), where $\Upsilon_{m}^{i}(\eta
,\lambda^{\lbrack l]})$ are some local functions. In particular, one can see
that%
\begin{equation}
\delta_{\nu}\lambda_{\chi}^{a}=\sum_{i=1}^{\aleph_{\chi}}\mathcal{D}^{ai}%
(\eta,\lambda^{\lbrack l]})\nu_{i}^{[i]}+O(\nu_{j}^{[l]}%
\,,\;l<j)\,.\label{sdc.22}%
\end{equation}

Note that the local functions $G_{m}^{i}$ , $\mathcal{C}_{ka}^{im}\,,$ and
$\Upsilon_{m}^{i}$ do not depend on the gauge parameters and are, in that
sense, universal. The matrices $\mathcal{C}$ and $\mathcal{D}$ are not singular.

\textbf{III.} \textbf{In theories with FCC, any symmetry }$\delta\mathbf{\eta
}$\textbf{, }$G$\textbf{\ of the Hamiltonian action }$S_{\mathrm{H}}%
$\textbf{\ can be represented as the sum of three types of symmetries}%
\begin{equation}
\left(
\begin{array}
[c]{c}%
\delta\mathbf{\eta}\\
G
\end{array}
\right)  =\left(
\begin{array}
[c]{c}%
\delta_{c}\mathbf{\eta}\\
G_{c}%
\end{array}
\right)  +\left(
\begin{array}
[c]{c}%
\delta_{\bar{\nu}}\mathbf{\eta}\\
G_{\bar{\nu}}%
\end{array}
\right)  +\left(
\begin{array}
[c]{c}%
\delta_{\mathrm{tr}}\mathbf{\eta}\\
G_{\mathrm{tr}}%
\end{array}
\right)  \,, \label{sdc.24}%
\end{equation}
\textbf{such that: }

\textbf{The set} $\delta_{c}\mathbf{\eta}$, $G_{c}$ \textbf{is} \textbf{a
global symmetry, canonical for the phase-space variables }$\eta$\textbf{. The
corresponding} \textbf{conserved charge }$G_{c}$\textbf{\ does not vanish on
the extremals.}

\textbf{The set} $\delta_{\bar{\nu}}\mathbf{\eta}$, $G_{\bar{\nu}}%
$\textbf{\ is a particular gauge transformation given by Eqs. }(\ref{sdc.5}),
(\ref{sdc.7}), \textbf{and} (\ref{sdc.8}) \textbf{with fixed gauge parameters
(i.e. with specific forms for the functions }$\nu_{i}=\bar{\nu}_{i}\left(
t,\eta^{\left[  l\right]  },\lambda^{\left[  l\right]  }\right)  $\textbf{)
that do not vanish on the extremals. The corresponding conserved charge
}$G_{\bar{\nu}}$ \textbf{vanishes on the extremals, whereas the variations
}$\delta_{\bar{\nu}}\mathbf{\eta}$ \textbf{do not.}

\textbf{The set} $\delta_{\mathrm{tr}}\mathbf{\eta}$, $G_{\mathrm{tr}}%
$\textbf{\ is a trivial symmetry. All the variations }$\delta_{\mathrm{tr}%
}\mathbf{\eta}$ \textbf{and the corresponding conserved charge}
$G_{\mathrm{tr}}$ \textbf{vanish on the extremals. The gauge charge
}$G_{\mathrm{tr}}$\textbf{\ depends on the extremals as }$G_{\mathrm{tr}%
}=O\left(  I^{2}\right)  .$

As an example, we consider a field model which includes a set of Yang-Mills
vector fields $\mathcal{A}_{\mu}^{a}$ , $a=1,...,r,$ and a set of spinor
fields $\psi^{\alpha}=\left(  \psi_{i}^{\alpha},\;i=1,...,4\right)  ,$%
\begin{align}
&  S=\int\mathcal{L}dx\,,\;\mathcal{L}=-\frac{1}{4}G_{\mu\nu}^{a}G^{\mu\nu
a}+i\bar{\psi}^{\alpha}\gamma^{\mu}\nabla_{\mu\beta}^{\alpha}\psi^{\beta
}-V(\psi,\bar{\psi})\,,\nonumber\\
&  G_{\mu\nu}^{a}=\partial_{\mu}\mathcal{A}_{\nu}^{a}-\partial_{\nu
}\mathcal{A}_{\mu}^{a}+f_{bc}^{a}\mathcal{A}_{\mu}^{b}\mathcal{A}_{\nu}%
^{c}\,,\;\;\nabla_{\mu\beta}^{\alpha}=\partial_{\mu}\delta_{\beta}^{\alpha
}-iT_{a\beta}^{\alpha}\mathcal{A}_{\mu}^{a}\,, \label{ex.1}%
\end{align}
where $V$ is the local polynomial in the field, which contains no derivatives.
The model is based on a certain global Lie group $G,$%
\begin{align*}
&  \psi\left(  x\right)  \overset{g}{\rightarrow}\exp\left(  i\nu^{\alpha
}T_{a}\right)  \psi\left(  x\right)  ,\;g\in G\,,\;\nu^{a},\;a=1,...,r,\\
&  T_{a}=T_{a}^{+}\,,\;\left[  T_{\alpha},T_{b}\right]  =if_{ab}^{c}%
T_{c}\,,\;\;f_{ab}^{k}f_{kc}^{n}+f_{bc}^{k}f_{ka}^{n}+f_{ca}^{k}f_{kb}%
^{n}=0\,.
\end{align*}

For $V=0,$ the action is invariant under gauge transformations ($\nu^{a}%
=\nu^{a}\left(  x\right)  $)%
\begin{equation}
\delta\mathcal{A}_{\mu}^{a}=D_{\mu b}^{a}\nu^{b}\,,\;\delta\psi=iT_{a}\psi
\nu^{a}\,,\;D_{\mu b}^{a}=\partial_{\mu}\delta_{b}^{a}+f_{cb}^{a}%
\mathcal{A}_{\mu}^{c}\,. \label{ex.2}%
\end{equation}
We assume the polynomial $V$ to be such that the whole action (\ref{ex.1}) is
invariant under the transformations (\ref{ex.2}) as well. Below we relate the
symmetry structure of the model with its constraint structure. To this end we
first reveal the constraint structure.

Proceeding to the Hamiltonian formulation, we introduce the momenta%
\[
p_{0a}=\frac{\partial\mathcal{L}}{\partial\mathcal{\dot{A}}^{0a}}%
=0\,,\;p_{ia}=\frac{\partial\mathcal{L}}{\partial\mathcal{\dot{A}}^{ia}%
}=G_{i0}^{a}\,,\;p_{\psi}=\frac{\partial_{r}\mathcal{L}}{\partial\dot{\psi}%
}=i\bar{\psi}\gamma^{0}\,,\;p_{\bar{\psi}}=\frac{\partial_{r}\mathcal{L}%
}{\partial\overset{\cdot}{\bar{\psi}}}=0\,.
\]
Thus, there exists a set of primary constraints$\;\Phi^{(1)}=\left(  \chi
_{a}^{(1)},\varphi_{\sigma}^{(1)},\sigma=1,2\right)  =0,$ where%
\[
\chi_{a}^{(1)}=p_{0a}\,,\;\varphi_{1}^{(1)}=p_{\psi}-i\bar{\psi}\gamma
^{0}\,,\;\varphi_{2}^{(1)}=p_{\bar{\psi}}\,.
\]
The total Hamiltonian reads $H^{\left(  1\right)  }=\int\mathcal{H}%
^{(1)}d\mathbf{x}$\textbf{,}%
\[
\mathcal{H}^{(1)}=\frac{1}{2}p_{ia}^{2}+\frac{1}{4}G_{ik}^{a2}-p_{\psi}%
\gamma^{0}\gamma^{k}\nabla_{k}\psi+\mathcal{A}^{0a}(D_{ib}^{a}p_{ib}-\bar
{\psi}\gamma^{0}T_{a}\psi)+V+\lambda_{\chi}^{a}\chi_{a}^{(1)}+\lambda
_{\varphi}^{\sigma}\varphi_{\sigma}^{(1)}\,.
\]
By performing the Dirac procedure, one can verify that there only appear
secondary constraints $\chi_{a}^{(2)}=0,$
\begin{align*}
&  \left\{  \varphi_{\sigma}^{(1)},H^{\left(  1\right)  }\right\}
=0\Longrightarrow\lambda_{\varphi}^{\sigma}=\bar{\lambda}_{\varphi}^{\sigma
}\left(  \mathcal{A},\psi,\bar{\psi}\right)  \,,\\
&  \left\{  \chi_{a}^{(1)},H^{\left(  1\right)  }\right\}  =0\Longrightarrow
\chi_{a}^{(2)}=D_{ib}^{a}p_{ib}+i\left(  p_{\psi}T_{a}\psi+p_{\bar{\psi}}%
\bar{T}_{a}\bar{\psi}\right)  \,,\;\left(  \bar{T}_{a}\right)  _{\beta
}^{\alpha}=-\gamma^{0}\left(  T_{a}^{\ast}\right)  _{\beta}^{\alpha}\gamma
^{0}\,.
\end{align*}
All the constraints $\varphi$ are second-class and all the $\chi$ are
first-class. It turns out that the complete set of constraints already forms
the orthogonal constraint basis, namely:%
\[
\varphi^{(1|1)}\equiv\varphi^{(1)},\;\chi^{(1|2)}\equiv\chi^{(1)}%
\,,\;\chi^{(2|2)}\equiv\chi^{(2)},
\]
and there are no constraints $\chi^{(1|1)},$%
\[%
\begin{array}
[c]{ccccc}%
\varphi^{(1|1)} & \rightarrow & \bar{\lambda} &  & \\
\chi^{(1|2)} & \rightarrow & \chi^{(2|2)} & \rightarrow & O\left(
\Phi\right)
\end{array}
.
\]
According to the general considerations, we chose the gauge charge in the
form
\[
G=\int\left[  \nu^{a}\chi_{a}^{(2|2)}+C^{a}\chi_{a}^{(1|2)}\right]
d\mathbf{x}\,,\;C^{a}=\left(  c_{b}^{a}\nu^{b}+d_{b}^{a}\dot{\nu}^{b}\right)
\,.
\]
Solving the symmetry equation (\ref{symeq}), we obtain $C^{a}=\dot{\nu}%
^{a}-\nu^{c}\mathcal{A}^{0b}f_{cb}^{a}=D_{0b}^{a}\nu^{b}\,.$ Thus,
\begin{align*}
&  G=\int\left[  p_{\mu a}D_{b}^{\mu a}\nu^{b}+i\left(  p_{\psi}T_{a}%
\psi+p_{\bar{\psi}}\bar{T}_{a}\bar{\psi}\right)  \nu^{a}\right]
d\mathbf{x\,,}\\
&  \delta\mathcal{A}_{\mu}^{a}=\left\{  \mathcal{A}_{\mu}^{a},G\right\}
=D_{\mu b}^{a}\nu^{b}\,,\;\delta\psi=\left\{  \psi,G\right\}  =iT_{a}\psi
\nu^{a}\,.\,
\end{align*}

\section{Main conclusions}

Below we summarize the main conclusions.

\textbf{Any symmetry transformation can be represented as a sum of three kinds
of symmetries: global, gauge, and trivial symmetries. The global part of a
symmetry does not vanish on the extremals, and the corresponding charge does
not vanish on the extremals as well. This separation is not unique. In
particular, the determination of the global charge from the corresponding
equation, and thus the determination of the global part of the symmetry is
then ambiguous. However, the ambiguity in the global part of a symmetry
transformation is always a sum of a gauge transformation and a trivial
transformation. The gauge part of a symmetry does not vanish on the extremals,
but the gauge charge vanishes on them. We stress that the gauge charge
necessarily contains a part that vanishes linearly in the FCC, and the
remaining part of the gauge charge vanishes quadratically on the extremals.
The trivial part of any symmetry vanishes on the extremals, and the
corresponding charge vanishes quadratically on the extremals.}

\textbf{The reduction of symmetry variations to extremals are global canonical
symmetries of the physical action, whose conserved charge is the reduction of
the complete conserved charge to the extremals.}

\textbf{Any global canonical symmetry of the physical action can be extended
to a nontrivial global symmetry of the complete Hamiltonian action.}

\textbf{There are no other gauge transformations that cannot be represented in
the form} (\ref{sdc.4}).

We stress that \textbf{in the general case the gauge charge cannot be
constructed with the help of any complete set of FCC only, for its
decomposition contains SCC as well}. A model for which the gauge charge must
be constructed both with the help of FCC and of SCC is considered in the
\emph{Example 1}.

Note that \textbf{in our procedure, generators (conserved charges) of
canonical and gauge symmetries may depend on Lagrange multipliers and their
time derivatives.} \textbf{This happens in the case when the number of stages
in the Dirac procedure is more than two.} In the \emph{Example 2} we represent
models that illustrate this fact.

\textbf{The gauge charge contains time derivatives of the gauge parameters
whenever there exist secondary FCC. Namely, the power of the highest time
derivative that enters the gauge charge is equal to }$\aleph_{\chi}%
-1,$\textbf{\ where }$\aleph_{\chi}$\textbf{\ is the number of the last stage
when new FCC still appear.} A simple model for which the gauge charge contains
a second-order time derivative of the gauge parameter is considered in the
\emph{Example 3.}

Since there is an isomorphism between symmetry classes of the Hamiltonian
action $S_{\mathrm{H}}$\ and the Lagrangian action\textbf{ }$S$\textbf{ the
symmetry structure of Lagrangian action }$S$\textbf{\ coincides with the
symmetry structure of the Hamiltonian action} $S_{\mathrm{H}}$, and is given
by all the assertions represented above. \textbf{As to the concrete form of a
symmetry transformation (symmetry transformation of the coordinates) of the
Lagrangian action }$S$\textbf{, it can be obtained as a reduction of the
symmetry transformation of the coordinates of the Hamiltonian action
}$S_{\mathrm{H}}$\textbf{\ by the substitution of all the Lagrange multipliers
and momenta via coordinates and velocities.}

{\LARGE Example 1: }Consider a Hamiltonian action $S_{\mathrm{H}}$ that
depends on the phase-space variables $(q_{i},p_{i},\;i=1,2,)$ and $(x_{a}%
,\pi_{a},\;\alpha=1,2),$ and of two Lagrange multipliers $\lambda_{\pi}$ and
$\lambda_{p}\,,$%
\begin{align*}
&  S_{\mathrm{H}}=\int\left[  p_{i}\dot{q}_{i}+\pi_{\alpha}\dot{x}_{\alpha
}-H^{(1)}\right]  dt\,,\;H^{(1)}=H_{0}^{(1)}+x_{1}q_{2}^{2}\,,\\
&  H_{0}^{(1)}=\frac{1}{2}\pi_{2}^{2}+x_{1}\pi_{2}+\frac{1}{2}p_{2}%
^{2}+\frac{1}{2}q_{2}^{2}+q_{1}p_{2}+\lambda_{\pi}\pi_{1}+\lambda_{p}p_{1}\,.
\end{align*}
One can see that the model has two primary constraints $\pi_{1}$ and $p_{1}.$
It is easy to verify that a complete set of constraints can be chosen as
$\chi=\left(  \pi_{1},\pi_{2}\right)  $ and $\varphi=\left(  q_{1},q_{2}%
,p_{1},p_{2}\right)  .$ Here $\chi$ are FCC and $\varphi$ are SCC. Thus, the
model is a gauge one. The peculiarity of the model is that gauge symmetries of
the action $S_{\mathrm{H}}$ have gauge charges which must be constructed with
the help of both FCC and SCC.

{\LARGE Example 2: }Consider a Hamiltonian action $S_{\mathrm{H}}$ that
depends on the phase-space variables $(q_{a},p_{a},\;a=1,2,)$ and $(x_{a}%
,\pi_{a},\;\alpha=1,2,3),$ and on a Lagrange multiplier $\lambda\,,$%
\begin{align*}
&  S_{\mathrm{H}}=\int\left[  p_{i}\dot{q}_{i}+\pi_{\alpha}\dot{x}_{\alpha
}-H^{(1)}\right]  dt\,,\;\;H^{(1)}=H_{0}^{\left(  1\right)  }+V\,,\;V=q_{1}%
x_{1}x_{3}^{2}\,,\\
&  H_{0}^{\left(  1\right)  }=\frac{1}{2}\left(  q_{i}^{2}+p_{i}^{2}\right)
+x_{1}\pi_{2}+x_{2}\pi_{3}+\frac{1}{2}x_{3}^{2}+\frac{1}{2}\pi_{2}%
^{2}+\frac{1}{2}\pi_{3}^{2}+\lambda\pi_{1}\,,
\end{align*}
The model has one primary constraint $\pi_{1}$. The peculiarity of the model
is that symmetries of the action $S_{\mathrm{H}}$ have charges that must
depend on Lagrange multipliers.

{\LARGE Example 3:} Consider a Lagrangian action that depends on the
coordinates $x,y,z,$%
\[
S=\frac{1}{2}\int\left[  \left(  \dot{x}-y\right)  ^{2}+\left(  \dot
{y}-z\right)  ^{2}\right]  dt\,.
\]
One can easily see that the action is gauge invariant under the following
transformations that include first and second-order time derivatives of the
gauge parameters,%
\[
\delta x=\nu\,,\;\delta y=\dot{\nu}\,,\;\delta z=\ddot{\nu}\,.
\]

\begin{acknowledgement}
Gitman is grateful to the Brazilian foundations FAPESP and CNPq for permanent
support; Tyutin thanks RFBR 02-02-16944, and LSS-1578.2003.2 for partial support.
\end{acknowledgement}


\begin{thebibliography}{9}                                                                                                %
\bibitem {Singular}P.M. Dirac, \emph{Lectures on Quantum Mechanics} (Yeshiva
University, New York 1964)

\bibitem {GitTy90}D.M. Gitman and I.V. Tyutin, \emph{Quantization of Fields
with Constraints} (Springer-Verlag, Berlin 1990)

\bibitem {HenTe92}M. Henneaux and C. Teitelboim, \emph{Quantization of Gauge
Systems} (Princeton University Press, Princeton 1992)

\bibitem {BorTy98}V.A. Borochov and I.V. Tyutin, Physics of Atomic Nuclei,
\textbf{61 }(1998) 1603; ibid \textbf{62} (1999) 1070

\bibitem {197}B. Geyer, D.M. Gitman, and I.V. Tyutin,\ J. Phys. A\textbf{36}
(2003) 6587

\bibitem {FulGiT}G. Fulop, D. Gitman, and I. Tyutin, Int. J.Theor. Phys.
\textbf{38} (1999) 1953

\bibitem {196}D. M. Gitman, and I.V. Tyutin,\ Gravitation \& Cosmology,
\textbf{8}, No.1-2 (2002) 138; \emph{Constraint reorganization consistent with
Dirac procedure,} Michael Marinov Memorial Volume: \emph{Multiple Facets of
Quantization and Supersymmetry}, ed. M. Olshanetsky and A. Vainstein (World
Publishing, Singapore 2002) pp.184-204
\end{thebibliography}
\end{document}